\documentclass[a4paper,twocolumn,prl,superscriptaddress]{revtex4}
\usepackage{amsmath,amssymb}
\usepackage{graphicx}
\usepackage{color}
\usepackage{subfigure}
\usepackage{hyperref}
\begin{document}
\title{Conformal symmetry of electron-hole puddles in ungated graphene}
\author{M. N. Najafi}
\affiliation{Department of Physics, University of Mohaghegh Ardabili, P.O. Box 179, Ardabil, Iran}
\author{M. Ghasemi Nezhadhaghighi}
\affiliation{Department of Physics, College of Science, Shiraz University, Shiraz 71454, Iran}
\date{\today}
%\end{comment}
\begin{abstract}
In this paper the mono-layer graphene at the charge neutrality point is considered whithin Thomas-Fermi-Dirac theory, treating inhomogeneous external potentials and electron-electron interactions on equal footing. We present some general considerations concerning the probability measure of the ground state charge density. The system shows degrees of self-similarity. By analyzing the ground state carrier density profile, we show that although it is not Gaussian, the critical exponents are consistent with Kondev hyper-scaling relations. Using Schramm-Loewner (SLE) evolution we show that the ungated graphene has conformal invariance and the random zero-charge density contours are SLE$_{\kappa}$ with $\kappa=1.8\pm 0.2$.
\end{abstract}
\maketitle
\begin{figure}[t]
\begin{center}
\includegraphics[width=45mm]{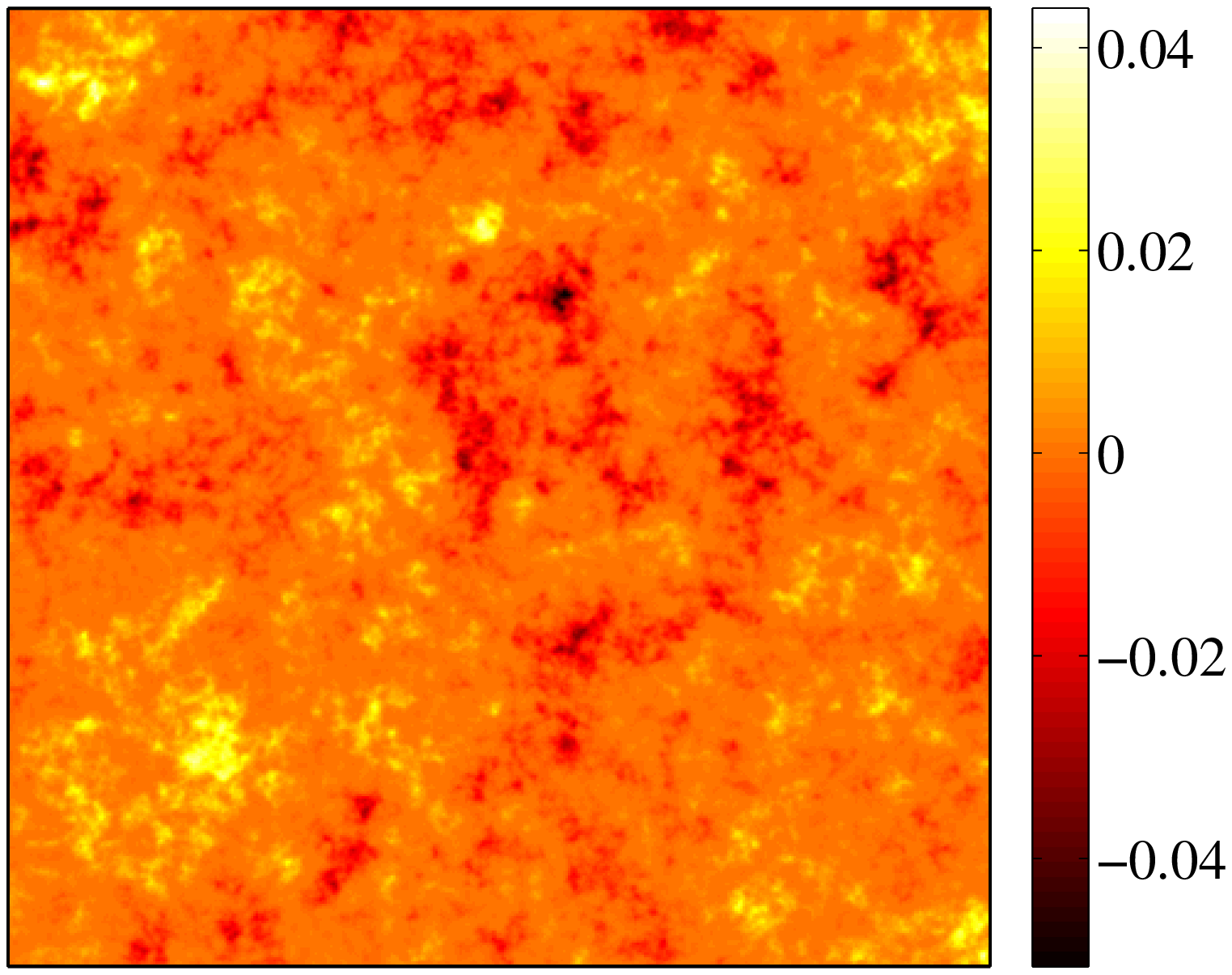}
\includegraphics[width=40mm]{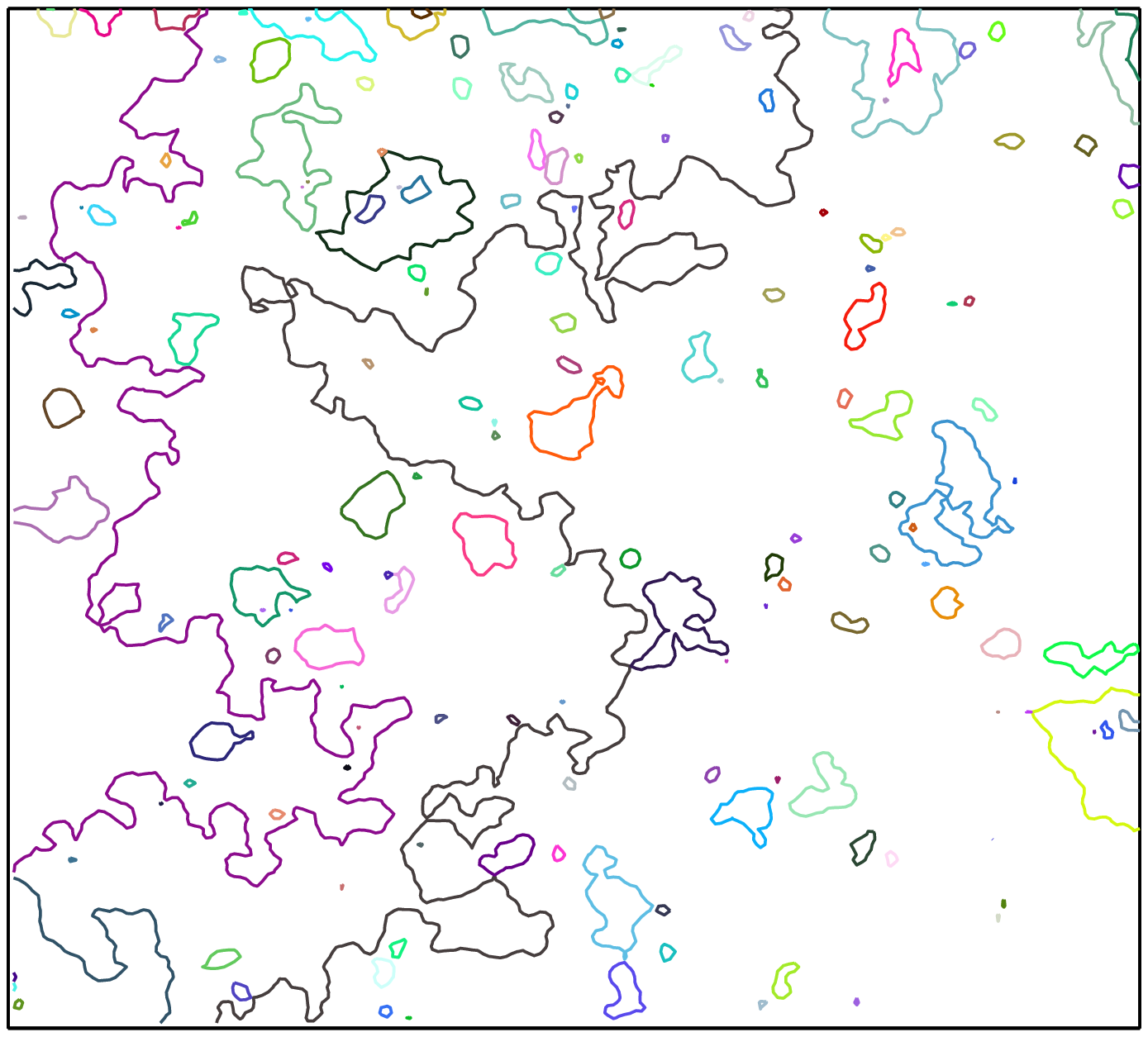}
\includegraphics[width=42mm]{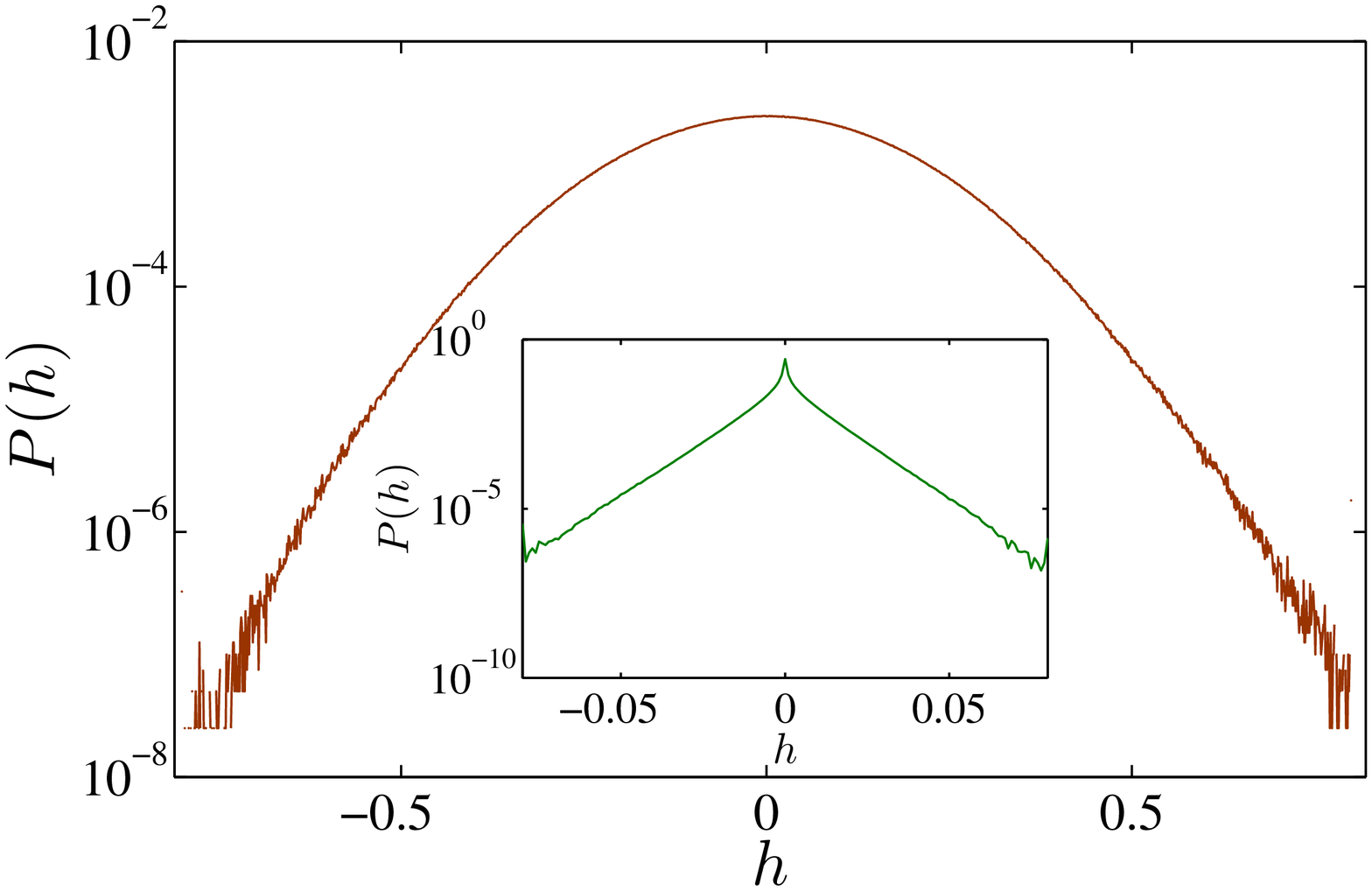}
\includegraphics[width=42mm]{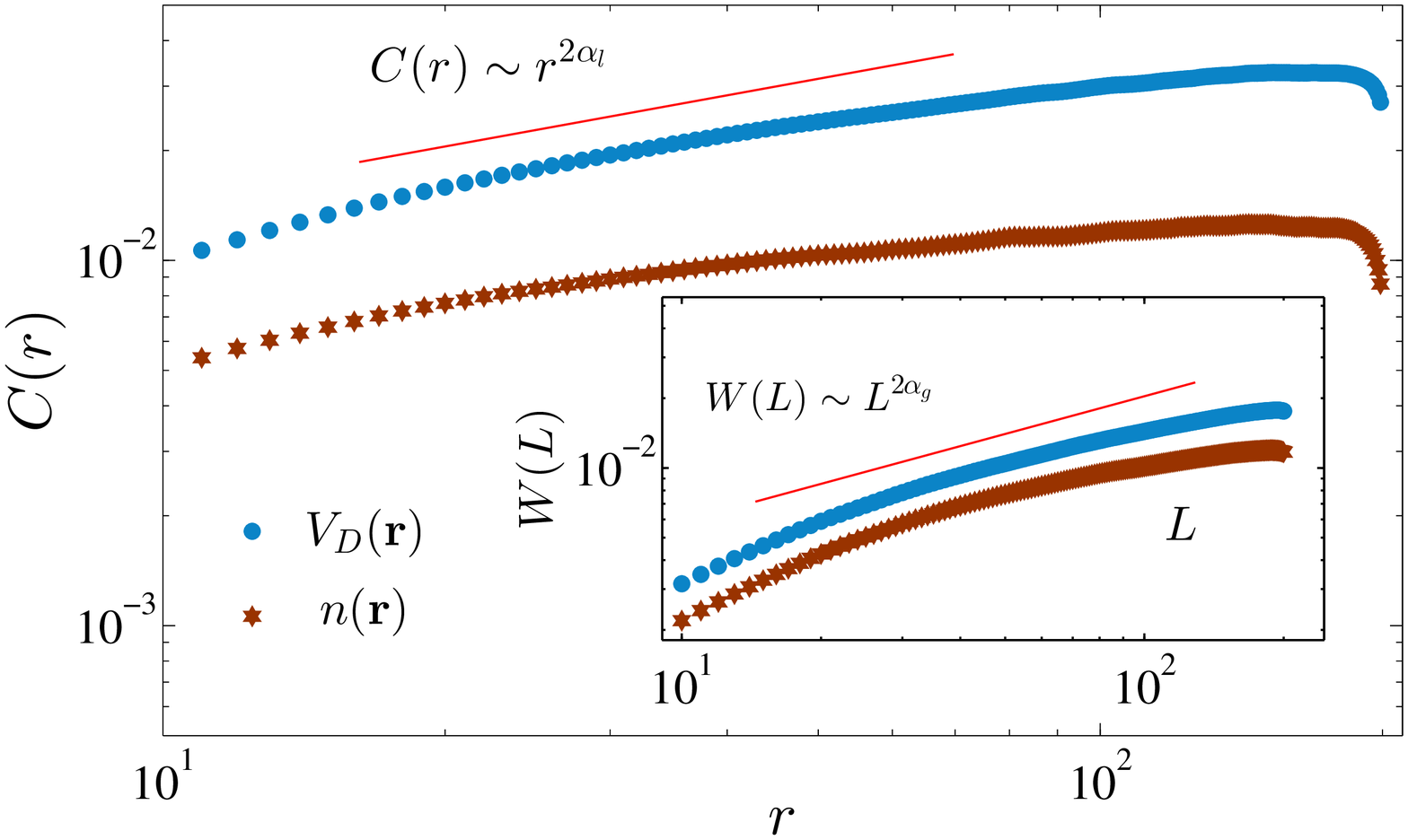}
\includegraphics[width=42mm]{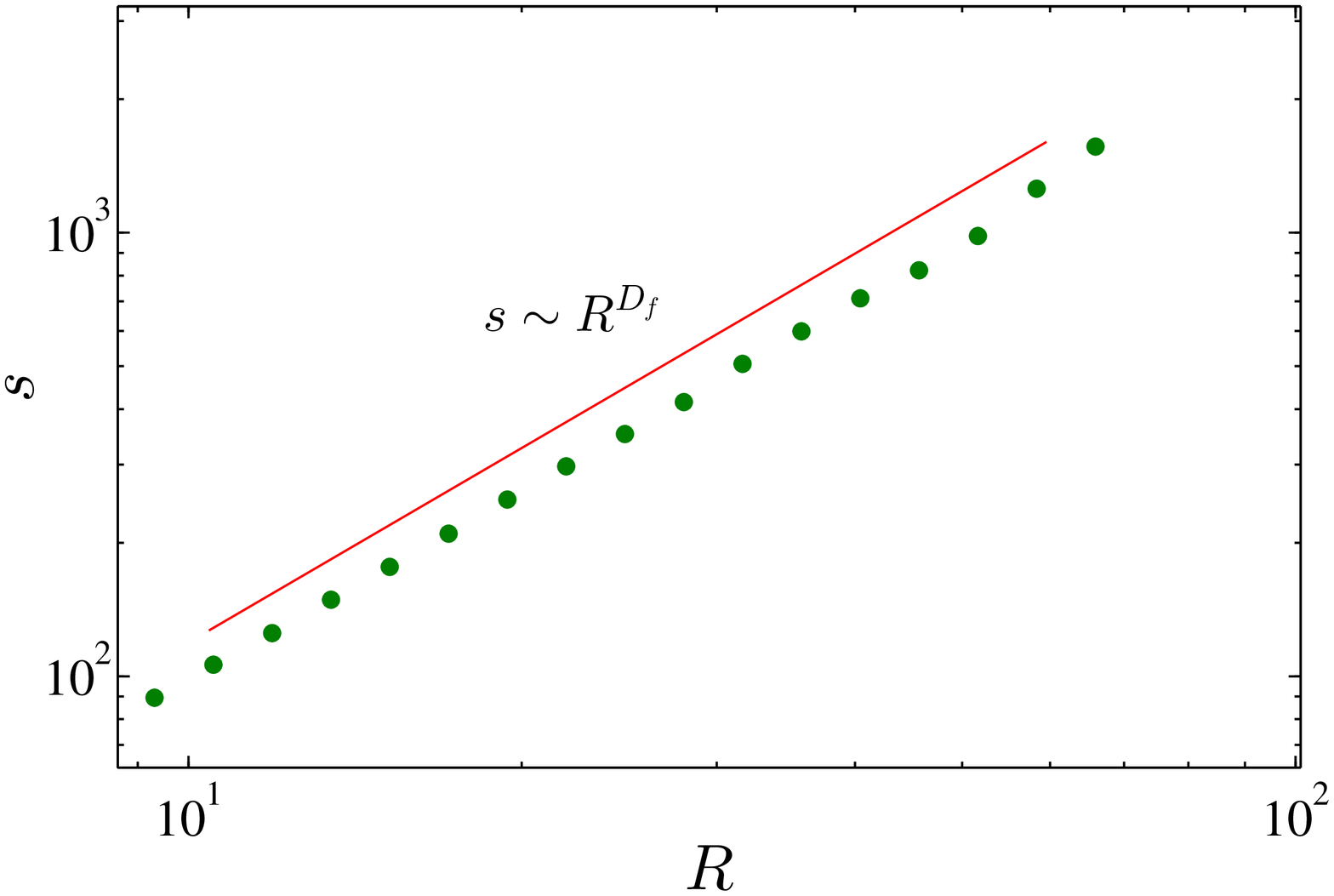}
\includegraphics[width=42mm]{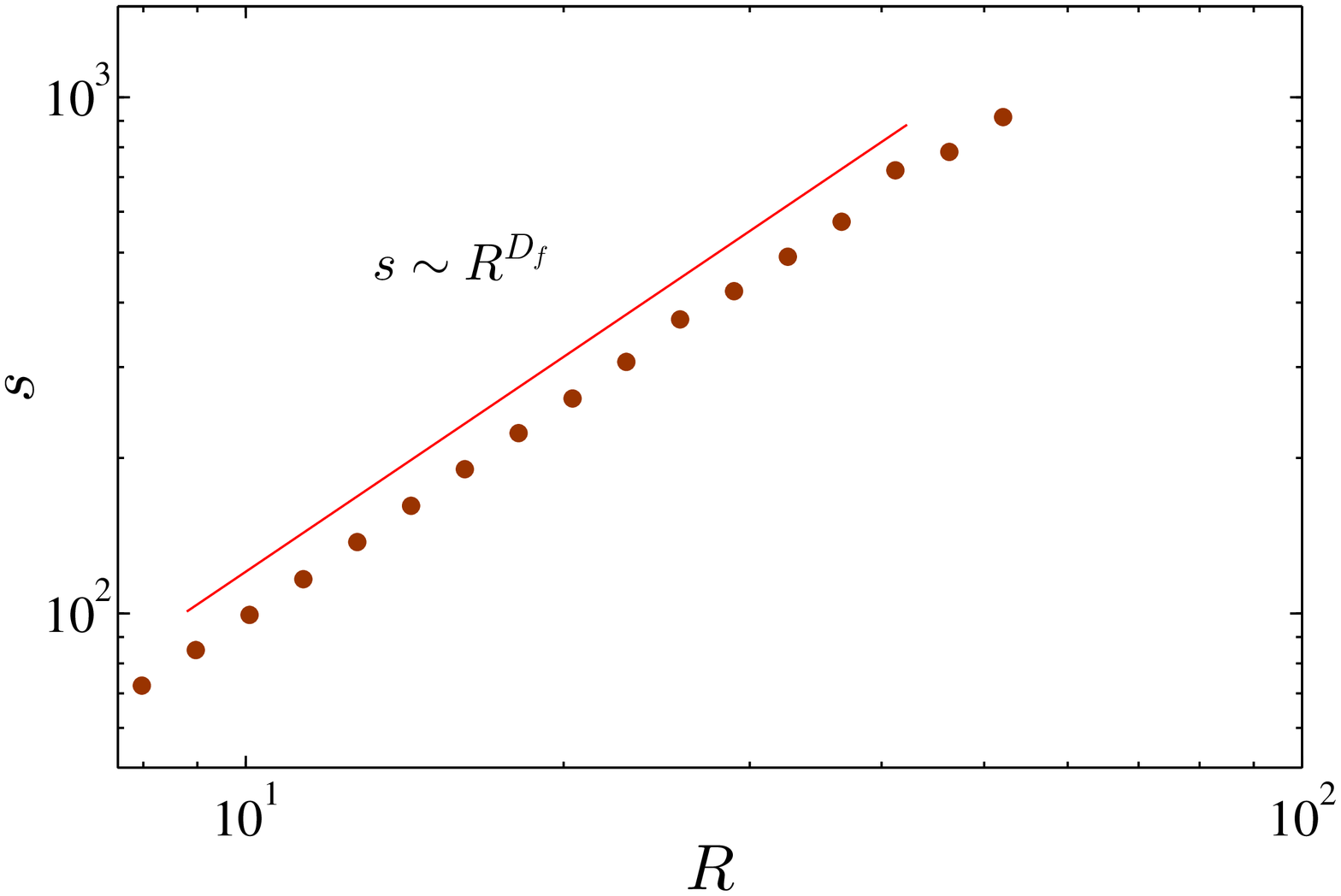}
\end{center}
\begin{picture}(100,0)(0,0)
\put(-50,240){(a)}
\put(80,240){(b)}
\put(-50,145){(c)}
\put(69,147){(d)}
\put(-50,80){(e)}
\put(75,80){(f)}
\end{picture}
\caption{(Color online) (a) A $400\ \text{nm}\times 400\ \text{nm}$ sample of carrier density and (b) its zero contour lines. (c) The probability measure of $n$ and $V_D$. (d) The charge and impurity potential correlation functions and total variance. The fractal dimension of loops for the (e) charge (f) impurity potential fields. The numerical values of exponents are: $\alpha_g^n=0.38\pm 0.02$, $\alpha_g^{V_D}=0.45\pm 0.02$, $\alpha_l^n=0.35\pm 0.03$, $\alpha_l^{V_D}=0.47\pm 0.02$, $D_f^{n}=1.39\pm 0.01$, $D_f^{V_D}=1.38\pm 0.02$}
\label{Figs1}
\end{figure}

Graphene is a two-dimensional system described by massless Dirac Fermions (MDF). The chiral nature of electrons in this system causes many interesting and strange features \cite{Polini}. The coexistence of disorder and particle-particle interaction in this system also leads to many interesting behaviors \cite{HwangAdamSarma}. Among these, the formation of electron-hole puddles (EHPs) in the low-densities is of special importance \cite{Rossi}. In sufficiently low densities in which EHPs form, the charge density fluctuations dominate the system and becomes larger than the average electron density in the system, driving the system into a new phase \cite{Rossi}. The saturation of conductivity in low densities (low-density minimal conductivity) is attributed to the formation of these EHPs. According to this idea the transport occure over the mentioned complex random network of conducting EHPs, leading to a saturation in conductivity. Despite a huge theoretical investigation on low density regime \cite{Polini,HwangAdamSarma,Rossi}, there is a little information concerning EHPs in graphene, espetially at the charge neutrality point (i.e. at the Dirac point) in which there are some indications of self-similarity and conformal symmetry \cite{Herrmann}.\\
EHPs were firstly predicted theoretically by Adam \textit{et al.} \cite{Adam} and Hwang \textit{et al.} \cite{HwangAdamSarma} and experimentally observed for mono-layer graphene (MLG) \cite{Martin,Rutter} and bi-layer graphene (BLG) \cite{Deshpande2}. For review see \cite{SarmaRevModPhys}. Among the experimental justifications of EHPs formation as high electron density inhomogeneity, the work of Martin \textit{et al.} is of espetial importance, since for the first time some statistics of EHPs were reported and the domainwalls of positive-density and negative-density clusters were drawn. This may be interpreted as the first attempt towards geometrical (global) approach to EHPs. After some statistical analysis, the typical spatial extension of EHPs were reported to be $\simeq 30$ nm (consistent with the theoretical results \cite{Rossi}) and the charge density fluctuations were calculated to be much more than the average carrier density over the graphene sheet. The other more detailed (direct and indirect) experiments also support these results. The other interesting experimental result, obtained by STM and SET experiments is that the rippling of graphene are independent of the charge density inhomogeneities, i.e. EHPs \cite{Rutter}. A substantial feature of experiments near the Dirac point is the formation of large (spanning) clusters of negative or positive charge densities. The presence of the spanning cluster in a system may be the fingerprint of a subtle symmetry; the scale invariance which leads to some scaling behaviors. If true, the system in hand lies within some universality class of the critical phenomena for which some non-perturbative techniques such as conformal field theory (CFT) and Schramm-Loewner evolution (SLE) should be employed.\\
Experiments, by themselves, are unable to directly identify the cause of the carrier density inhomogeneities (EHPs) and characterize them. Despite of many successes in predicting transport properties in the presence of charge impurities (which is the main source of disorder in low carrier densities \cite{Barlas,Rossi}) the origin of EHPs and their physical properties have been poorly understood \cite{Adam,Nomura,Rossi}. The main trouble arise from the simultaneous vital role of disorder and interaction. The marginal character of interaction in graphene leads to many interesting properties of graphene \cite{Gonzalez,HwangSarma,Vafek,Polini,Barlas}, as well as the peculiar dependence of the exchange-correlation energy to the charge density which is the source of many differences of graphene from the other systems \cite{Barlas}. It has been shown that in the low density limit the exchange-correlation potential is $V_{xc}=\frac{1}{4}\left[ 1-gr_s\zeta(gr_s)\right]\text{sgn}(n)\sqrt{\pi|n|}\ln\left(4k_c/\sqrt{4\pi |n|}\right)$ (in which $k_c$ is the momentum cut-off and $\zeta(y)=\frac{1}{2}\int_0^{\infty}\frac{dx}{(1+x^2)^2\left( \sqrt{1+x^2}+\pi y/8\right)}$) which is completely different from ordinary 2D parabolic band systems \cite{Barlas}. In the Thomas-Fermi-Dirac theory employed in this paper, we use this dependence.\\
The role of various disorders in graphene have been largely investigted \cite{Nomura,HwangAdamSarma}. The approximately linear dependence of conductivity on carrier density in graphene sheets indicates that the remote charge impurities are dominant disorder source in most graphene samples, which locally shifts the Dirac point \cite{HwangAdamSarma}. In vicinity of the charge neutrality (Dirac) point, the screening is low, implying that the Coulomb impurities with the potential should be taken into account. As an important attempt to bring the effect of interaction and correlations simultaneously in the problem, DFT-LDA approach \cite{Polini} has serious limitations, namely numerical complexity and smallness of samples. \\
The case of interest in this paper, which makes the problem more tractable and treats the interaction and disorder on equal footing is an slow (spatial) varying charge density system for which the Thomas-Fermi-Dirac theory is applicable. Using the local density approximation one can prove that the total energy of the graphene is \cite{SarmaRevModPhys}:
\begin{eqnarray}
\begin{split}
E=&\hbar v_F[\frac{2\sqrt{\pi}}{3}\int d^2\textbf{r}\ \text{sgn}(n(\textbf{r}))|n(\textbf{r})|^{\frac{3}{2}}\\
&+\frac{r_s}{2}\int d^2\textbf{r}\int d^2\textbf{r}^{\prime}\frac{n(\textbf{r})n(\textbf{r}^{\prime})}{|\textbf{r}-\textbf{r}^{\prime}|}\\
&+r_s\int d^2\textbf{r}V_{xc}[n(\textbf{r})]n(\textbf{r})+r_s\int d^2\textbf{r}V_D(\textbf{r})n(\textbf{r})\\
&-\frac{\mu}{\hbar v_F}\int d^2\textbf{r}n(\textbf{r})]
\end{split}
\end{eqnarray}
in which $v_F$ is the Fermi velocity, $r_s\equiv e^2/\hbar v_F\kappa_S$ is the dimensionless interaction coupling constant, $\mu$ is the chemical potential, $g=g_sg_v=4$ is the total spin and valley degeneracy. The remote Coulomb disorder potential is $V_D(\textbf{r})=\int d^2\textbf{r}^{\prime}\frac{\rho(\textbf{r}^{\prime})}{\sqrt{|\textbf{r}-\textbf{r}^{\prime}|^2+d^2}}$ in which $\rho(r)$ is the charged impurity density and $d$ is the distance between substrate and the graphene sheet. In the above equations bare coulomb interactions were taken into account. By minimizing the energy we obtain:
\begin{eqnarray}
\begin{split}
&\text{sgn}(n(\textbf{r}))\sqrt{|\pi n(\textbf{r})|}+\frac{r_s}{2}\int d^2 \textbf{r}^{\prime}\frac{n(\textbf{r}^{\prime})}{|\textbf{r}-\textbf{r}^{\prime}|}\\
&+r_sV_{xc}[n(\textbf{r})]+r_sV_D(\textbf{r})-\frac{\mu}{\hbar v_F}=0.
\end{split}
\label{mainEQ}
\end{eqnarray}
The disorder is assumed to be white noise with Gaussian distribution $\left\langle \rho(\textbf{r})\right\rangle=0 $ and $\left\langle \rho(\textbf{r})\rho(\textbf{r}^{\prime})\right\rangle=(n_id)^2\delta^2(\textbf{r}-\textbf{r}^{\prime})$. For the graphene on the SiO$_2$ substrate (to be used in this paper), the parameters are: $\kappa_S\simeq 2.5$, so that $r_s\simeq0.8$, $d\simeq 1$ nm, $k_c=1/a_0$ where $a_0$ is the graphene lattice constant $a_0\simeq 0.246$ nm corresponding to energy cut-off $E_c\simeq 3$ eV.\\
Equation (\ref{mainEQ}) has interesting scaling properties at $\mu=0$. To see this let us make the transformation $\textbf{r}\rightarrow \lambda \textbf{r}$. In the absence of $V_{xc}$, if we transform $n(\textbf{r})\rightarrow n(\lambda\textbf{r})=\lambda^{-2}n(\textbf{r})$ noting that $V_D(\lambda\textbf{r})=\lambda^{-1}V_D(\textbf{r})$, the Eq. \ref{mainEQ} remains unchanged, signaling the scaling behaviors of the surface. When $V_{xc}$ is included the Eq. \ref{mainEQ} remains unchanged provided that the coefficient of the first term of this equation becomes $1-\frac{1}{4}\left( 1-gr_s\zeta (gr_s)\right) r_s\ln\lambda$. Therefore the first term survives marginally in the infra-red limit. This scale-invariance in two dimensions may lead to some power-law behaviors and some exponents which are vital for surface characterization. It may also lead to conformal invariance as a $2+1$ dimensional system, which determines its universality class. The above symmetry is simply an additional symmetry which limits the correlation functions to show power-law behaviors, but further details of the system needs analytical or numerical solution. One of the most important quantities in random field analysis is the probability measure of charge density $P(n)$ which is believed to be non-Gaussian in the case of charge density in graphene \cite{SarmaRevModPhys}. In the followings we use the fact that $\text{d}n=-r_sF_n\left[\text{d}\chi_{\rho}(d)+\frac{1}{2}\text{d}\chi_n(0)\right] $ in which $F_n\equiv\frac{2\text{sgn}(n)\sqrt{|n|/\pi}}{1-r_s\beta\left[\text{sgn}(n)-\ln\frac{4k_c}{\sqrt{4\pi |n|}}\right]}$ and $\text{d}\chi_x(d)\equiv\int d^2\textbf{r}^{\prime}\text{d}\left[x(\textbf{r}^{\prime})\right] \left(|\textbf{r}-\textbf{r}^{\prime}|^2+d^2\right)^{-1/2}$ and $x=\rho , n$. These equations are obtained directly by using Eq. \ref{mainEQ} and some staraightforward calculations. By some lengthy Ito calculations and using the homogeneity of the system we reach the following formula for the probability measure of the charge density $P\left( \left\lbrace n\right\rbrace \right) $
\begin{eqnarray}
\partial_nP(\left\lbrace n\right\rbrace ) =- \Gamma_n P(\left\lbrace n\right\rbrace )
\label{MainPnEq}
\end{eqnarray}
in which $\Gamma_n\equiv\zeta_0\frac{G_n}{F_n}+2\partial_n\ln F_n $, $\zeta_0\equiv \frac{2}{\pi dn_i^2r_s}$, $G_n=\int d^2\textbf{r}^{\prime}\frac{n(\textbf{r}^{\prime})}{|\textbf{r}-\textbf{r}^{\prime}|^2}$, and $\partial_n$ is the functional derivative. This equation is the master equation governing the probability distribution of a density configuration. For the local charge probability distribution $P(n)$ it is sufficient to use the independence of $P(n)$ of the spatial point $\textbf{r}$. The result is the same as Eq. \ref{MainPnEq}, replacing $P_n(\left\lbrace n\right\rbrace )$ simply by $P(n)$ and the functional derivative by simple derivative, i.e. $\partial_nP(n) =- \Gamma_n P(n)$. Let us now look at the weak coupling limit $r_s\rightarrow 0$, or the weak disorder limit $n_i\rightarrow0$, i.e. large $\zeta$ limit. To facilitate the calculations let us also assume that $G_n$ is nearly constant from which we have $G_n\varpropto \left\langle n\right\rangle$. This yields $\Gamma_n\simeq \frac{\sqrt{\pi}}{2}\zeta_0^{\prime}\frac{\text{sgn}(n)}{\sqrt{|n|}}=\zeta_0^{\prime}\partial_n\left(\text{sgn}(n)\sqrt{\pi|n|}\right)$ in which $\zeta_0^{\prime}\equiv \zeta_0 G_n$. The solution for local probability distribution is therefore $P_n=A\exp\left[-\zeta^{\prime}\left(\text{sgn}(n)\sqrt{\pi |n|}-\frac{\mu}{\hbar Sv_F}\right)\right]$ in which $A$ is a normalization constant and $S$ is the area of the sample. Due to violating the particle-hole symmetry in the case $\mu=0$, this relation may seem not to be correct. To answer, we should consider the approximated $G_n$ whose amount grows negatively for negative $n$ values. In fact it restores the electron-hole symmetry, resulting to an electron-hole symmetric $P_n$. In this equation, the effects of disorder and Hartree interaction have been coded in $\zeta^{\prime}$ which diverges for very weak or disorder, resulting to a very wide charge distribution and large charge fluctuations. Note also that the $\mu\rightarrow 0$ limit has direct effect on $G_n$, i.e. $\mu$ controls $\left\langle n\right\rangle$ and consequently $G_n$. Therefore as $\mu\rightarrow 0$, $G_n$ is expected to become vanishingly small and ${\zeta^{\prime}}^{-1}\rightarrow \infty$. We see that at the charge neutrality point the density fluctuations grow unboundedly which implies the formation of large scale inhomogeneities for which the power-law behaviors become possible.\\
To investigate the properties of the system for arbitrary strengths of coupling and disorder we solved Eq. \ref{mainEQ} numerically. The steepest descent method was used to solve Eq. \ref{mainEQ} iteratively. In our numerical process, we discretized the real space by $1$ nm steps and generated $L\times L$ square lattice. We repeated our analysis for $L=50$ nm, $100$ nm, $200$ nm, $300$ nm and $400$ nm to control the finite size effects. We found that the results are independent of the system size for $L\gtrsim 100$ nm. Over $6\times 10^3$ samples for each system size were generated (the total (2.4 GHz) CPU time spent was $1.2\times 10^8$ s). The self-consistency parameter was set to $10^{-10}$. A charge sample and its zero-contours has been shown in Fig. \ref{Figs1}a and \ref{Figs1}b. From the Fig. \ref{Figs1}b we see that the probability distribution of impurity potential $P(V_D)$ is Gaussian as expected, whereas $P(n)$ is not in agreement with the other predictions \cite{SarmaRevModPhys}. Other statistical tests also support this result. We have also calculated some other exponents of the system, supporting the hypothesis that the ungated samples are self-similar. It is well-known that for a scale-invariant surface the correlation function $C(r) \equiv \langle \left[ X(\mathbf{r}+\mathbf{r_0})-X(\mathbf{r_0}) \right]^2 \rangle $ shows power-law bahaviors, i.e. $C(r)\sim |\mathbf{r}| ^{2\alpha_l}$ in which $\alpha_l$ is called the local roughness exponent and $X=n,V_D$. The other important quantity is the global roughness exponent for scale-invariant rough surfaces defined by $W(L)\equiv \langle \left[ X(\mathbf{r}) - \bar{X} \right]^2 \rangle _L \sim L^{2\alpha_g}$ where $\bar{X}=\langle X(\mathbf{r}) \rangle_L$, and $\langle \dots \rangle _L$ means that, the average is taken over $\mathbf{r}$ in a box of size $L$. For Gaussian surfaces, $\alpha_l=\alpha_g$. These exponents have been shown in Figs. \ref{Figs1}d from which the power-law behavior is evident. The numerical values have been also reported in its caption. The important eponent which directly show the geometrical properties of the model in hand is the fractal dimension ($D_F$) of loops. This is defined by $\left\langle \log l\right\rangle=D_F \left\langle \log r\right\rangle $ ($l$ is are the loop length and $r$ is the gyration radius of loop and $\left\langle \right\rangle$ is the ensemble average) and may directly reflect the conformal symmetry of the system \cite{Cardy}. These are shown separately for charge density and impurity potential in Fig. \ref{Figs1}e and \ref{Figs1}f, indicating that the fractal dimesions are the same within their error bars. Their numerical values are not far from the one for the domainwalls of spin clusters of 2D critical Ising model \cite{NajafiSLEkr2}. We have observed that, although not a Gaussian random surface, $n$ satisfied the Kondev hyper-scaling relations \cite{Kondev}.\\

\begin{figure}[t]
\begin{center}
\includegraphics[width=95mm]{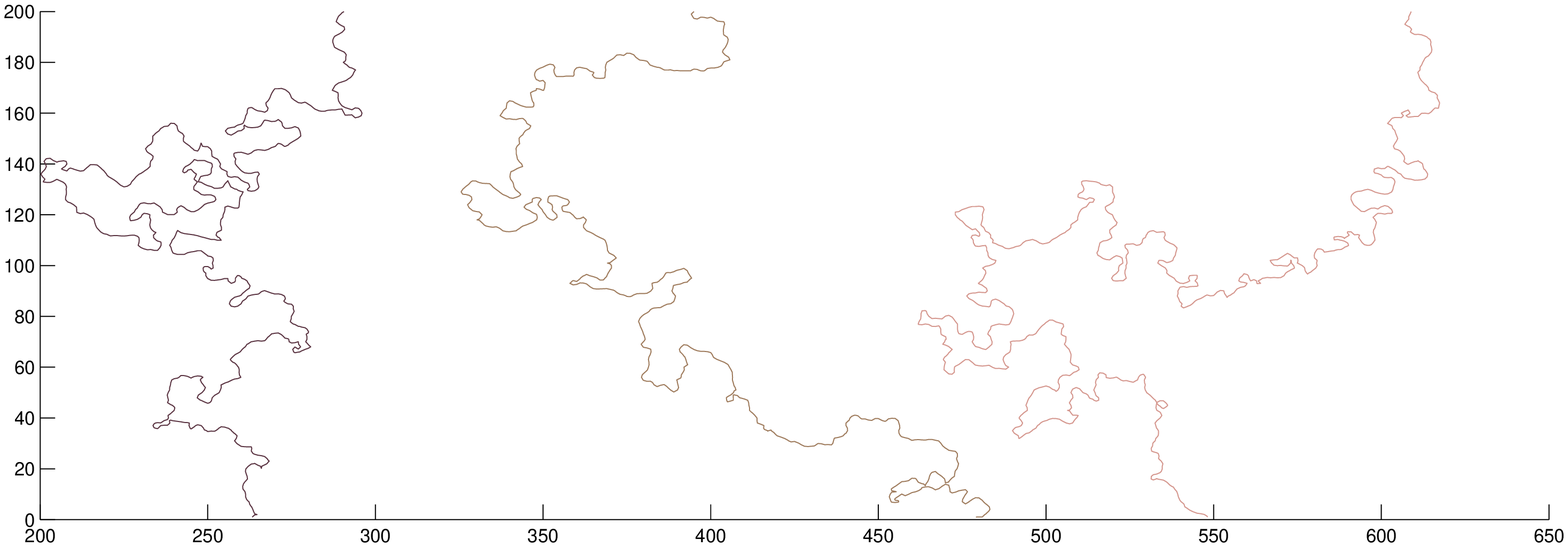}
\includegraphics[width=42mm]{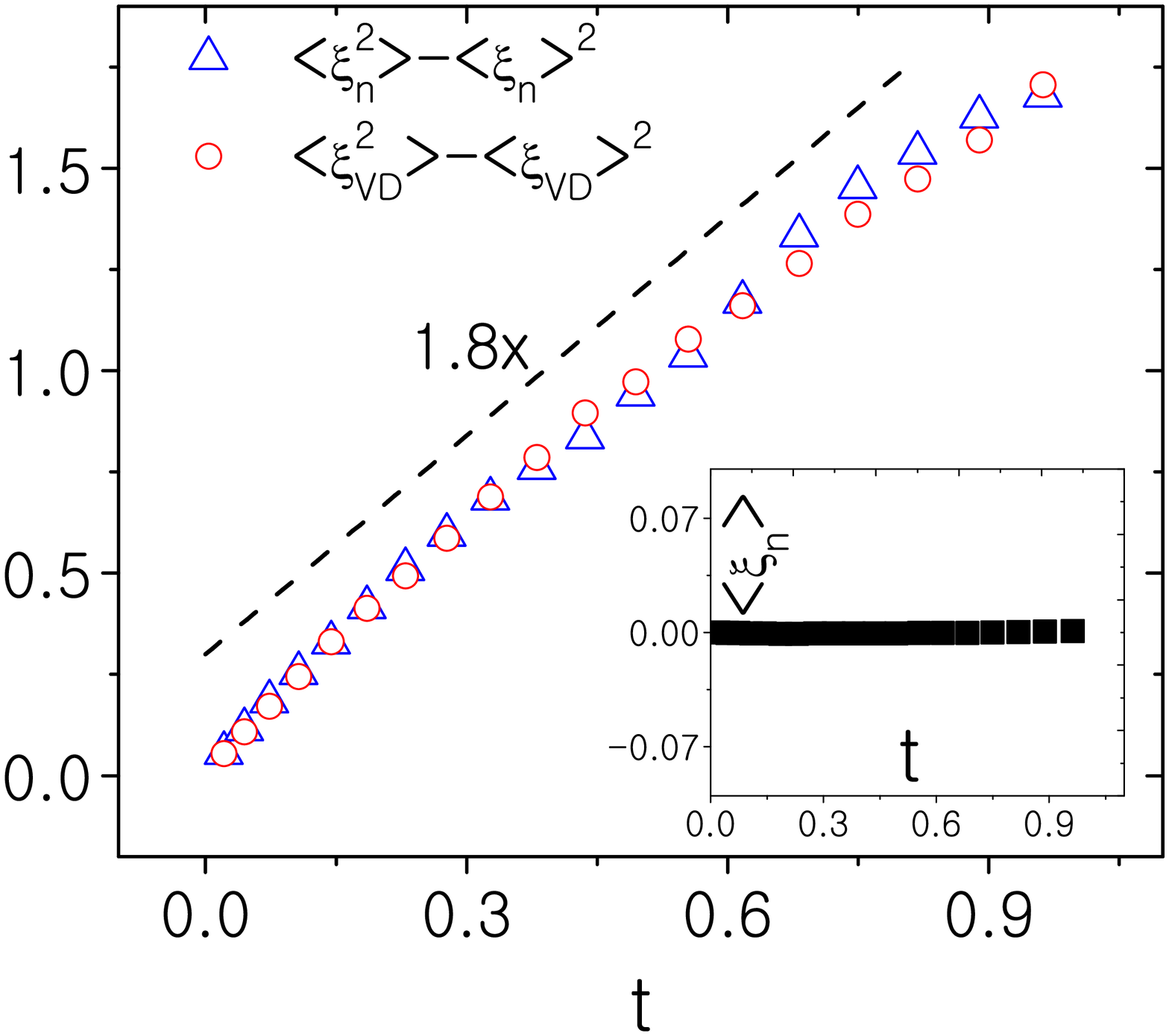}
\includegraphics[width=42mm]{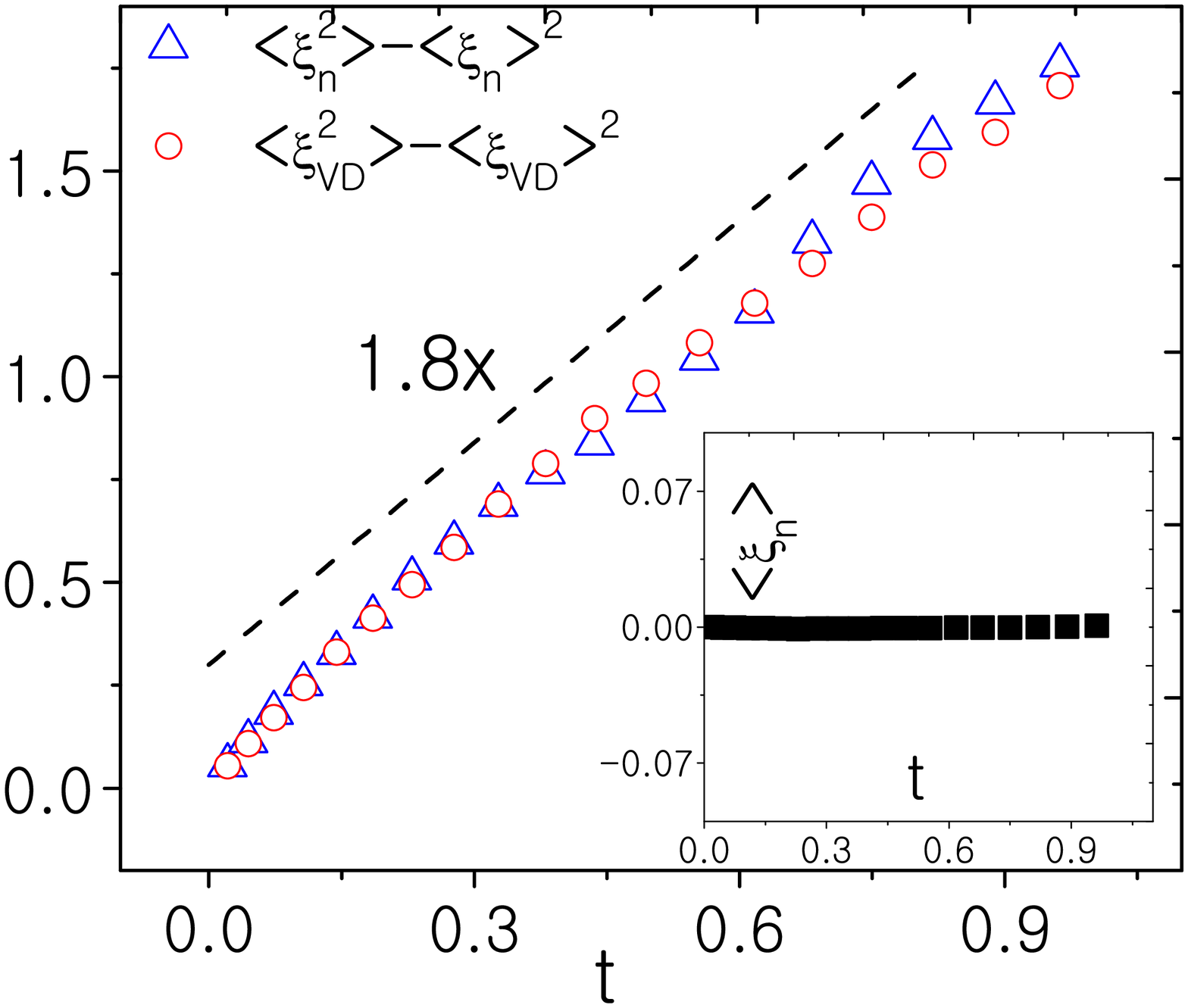}
\end{center}
\begin{picture}(100,0)(0,0)
\put(0,170){(a)}
\put(-20,10){(b)}
\put(100,10){(c)}
\end{picture}
\caption{(Color online) (a) Some interfaces linking the lower boundary to the upper one. $\left\langle \xi^2\right\rangle-\left\langle \xi\right\rangle^2 $ as a function of $t$ for charge density $n$ and impurity potetial $V_D$, analyzed by (b) slit map (c) strip map. The inset graphs show that $\left\langle \xi\right\rangle\simeq 0$.}
\label{Figs2}
\end{figure}

\textbf{SLE investigation}: According to SLE theory one can describe the geometrical objects (which may be interfaces) of a 2D critical model via a growth process and classify them into one parameter ($\kappa$) classes \cite{Cardy}. From a simple relation between the central charge $c$ in conformal field theory (CFT) and the diffusivity parameter $\kappa$ in SLE, namely $c=\frac{(6-\kappa)(3\kappa-8)}{2\kappa}$, one can find the corresponding CFT \cite{Cardy,NajafiPRE1,NajafiPRE2,BauBer}, and consequently the universality class is obtained. Chordal SLE$_{\kappa}$ is a growth process defined via conformal maps, $g_{t}(z)$, which are solutions of the Loewner's equation $\partial_{t}g_{t}(z)=\frac{2}{g_{t}(z)-\xi_{t}}$ where the initial condition is $g_{t}(z)=z$ and $\xi_{t}$ (the driving function) is a continuous real valued function which is shown to be proportional to the one dimensional Brownian motion ($\xi_t=\sqrt{\kappa}B_t$) if the curves have two properties: conformal invariance and the domain Markov property. If one has an ensemble of conformal loops, should use a further analysis on loops, since the chordal SLE describes the curves going from origin to infinity. In this case one can take the following steps to extract $\kappa$ \cite{NajafiSLEkr2}: (I) Cut the loops horizontally and then send its end point to the infinity by the map $\phi(z)=\frac{x_{\infty}z}{z-x_{\infty}}$ in which $x_{\infty}$ is the end point of the cut curve and $z=x+iy$ is the complex coordinate in the upper half plane. (II) Assume the driving function to be partially constant in each time interval and discretize the Loewner's equation. (III) Uniformize the curve step by step and in each time step $t$, set $\xi_t$ equal to mapped point of the tip of the curve at that time. (IV) Verify that $\left\langle\xi_t\right\rangle=0$ and calculate the slope of $\left\langle\xi_t^2\right\rangle-\left\langle\xi_t\right\rangle^2$ versus time $t$, i. e. $\left\langle\xi_t^2\right\rangle-\left\langle\xi_t\right\rangle^2=\kappa t$ which yields the diffusivity parameter $\kappa$.\\
In our case, the loops are not so large to use this algorithm for which the error bar of $\kappa$ is large. Therefore we carried out another simulation (like above simulation in sprit) of lattice size $400\times 400$ and extracted the zero-charge lines \textit{going from one boundary to the opposite one}. Some samples of this kind have been shown in Fig. \ref{Figs2}a. For extracting the diffusivity parameter, we have used two SLE methods to be more precise: conformal map on the upper half plane and on the strip geometry. For the former we have used the slit map \cite{NajafiSLEkr,NajafiWater}, whereas for the later we have used strip map \cite{NajafiTavana}. For the review on SLE maps see \cite{BauBer}. The results of the upper half plane have been shown in Fig. \ref{Figs2}b and for the strip geometry in Fig. \ref{Figs2}c. It is notable that both vertical and horizontal axes have been re-scaled to unity. In the inset graph of these figures $\left\langle \xi_n\right\rangle $ and $\left\langle \xi_{V_D}\right\rangle$ have been sketched and it is seen that they are nearly zero for both slit and strip cases as required for SLE curves. The slopes for both cases coincide with each other showing that our analysis is independent of the method. Interestingly it is seen that the diffusivity parameter for both $n$ and $V_D$ are the same within their error bars, i.e. $\kappa_{n}=\kappa_{V_D}=1.8\pm0.2$ for both slit and strip maps. For minimal conformal models $\kappa$ should be in the interval $\left[ 2,8\right] $. The fact that our result does not lie whithin this interval implies that it does not match with any minimal model. The $\kappa$ less than 2 has recently observed also for the watershed of random landscapes \cite{daryaii}. It seems peculiar that a Gaussian and non-Gaussian random fields have the same critical exponents and diffusivity parameter. We present an analysis concerning this point. According to Hohenberg-Kohn theorem there is a one to one correspondence between the ground state charge density of a quantum system (here $n(\textbf{r})$) and the external potential (here $V_D$). This can be expressed by the relation $V_D=V_D[n]$ which may be a non-local function. Therefore the characteristic level lines of $V_D$ results in the same level lines for $n$ and the statistics are similar. Now consider the probability measure of them, i.e. $P(V_D)$ and $P(n)$. The equality of the probabilities implies that $P(V_D)\text{d}V_D=P(n)\text{d}n$, according to which we have $P(n)=\left( \text{d}V_D/\text{d}n\right) P(V_D)$. Note that the necessary condition for this relation is that the conditional probability function $P(n|V_D)$ be a narrow function of both $V_D$ and $n$. This implies that, given that $P(V_D)$ is Gaussian, the function $P(n)$ may not, depending on the quantity $\text{d}V_D/\text{d}n$.\\
We conclude that the ungated graphene is an scale-invariant 2D system with peculiar charge density profile. Although it is not Gaussian, it shows critical behaviors for which the critical exponents satisfy the hyper-scaling relations of Kondev. We observed that the charge density domainwalls, when treated as stochastic curves respect the SLE requirements ($\kappa\simeq1.8$) reflecting the fact that the system in hand has conformal symmetry.

\end{document}